\documentclass[aps,twocolumn,showpacs,byrevtex,prl,reprint,nofootinbib]{revtex4-1}
\usepackage{xspace}
\usepackage{epsfig}
\usepackage{dcolumn}
\usepackage{amssymb}   % for math
\usepackage{amsmath}
\usepackage{mathrsfs}
\usepackage{ulem}
\usepackage{bm}
\usepackage{color}
\usepackage{graphicx}
\usepackage{subfigure}
\usepackage{pstricks}
\usepackage{pst-node}
\usepackage{rotating}
\usepackage{longtable}
\usepackage{appendix}
\usepackage{flushend}

\usepackage{times}
\usepackage[english]{babel}
\addto{\captionsenglish}{%

}
\usepackage[colorlinks=true,linkcolor=blue,filecolor=blue,urlcolor=blue,citecolor=blue]{hyperref}

\usepackage{lineno}
\usepackage{dcolumn}
\newcolumntype{d}[1]{D{.}{.}{#1}}

\newcommand{\EE}{e^+e^-}

\newcommand{\ar}{\rightarrow}

\newcommand{\bfg}{\begin{figure}}
\newcommand{\efg}{\end{figure}}
\newcommand{\bitm}{\begin{itemize}}
\newcommand{\eitm}{\end{itemize}}
\newcommand{\bnum}{\begin{enumerate}}
\newcommand{\enum}{\end{enumerate}}
\newcommand{\btbl}{\begin{table*}}
\newcommand{\etbl}{\end{table*}}
\newcommand{\btbu}{\begin{tabular}}
\newcommand{\etbu}{\end{tabular}}
\newcommand{\bcl}{\begin{center}}
\newcommand{\ecl}{\end{center}}
\newcommand{\bbt}{\bibitem}
\newcommand{\beq}{\begin{equation}}
\newcommand{\eeq}{\end{equation}}
\newcommand{\beqr}{\begin{eqnarray}}
\newcommand{\eeqr}{\end{eqnarray}}

\begin{document}
\normalsize
\parskip=5pt plus 1pt minus 1pt
\title{\boldmath Precise Measurement of Born Cross Sections for $\EE\ar D\bar{D}$ at $\sqrt{s} = 3.80-4.95$ GeV
}
\author{BESIII Collaboration
\\ 
~[Published in Phys. Rev. Lett. {\bf 133}, 081901 (2024)]
}
%\date{\today}
\thanks{Full author list given at the end of the Letter}
\begin{abstract}
Using data samples collected with the BESIII detector at the BEPCII collider at center-of-mass energies ranging from 3.80 to 4.95 GeV, corresponding to an integrated luminosity of 20 fb$^{-1}$, a measurement of Born cross sections for the $\EE\ar D^{0}\bar{D}^{0}$ and $D^{+}D^{-}$ processes is presented with unprecedented precision.  
Many clear peaks in the line shape of $\EE\ar D^{0}\bar{D}^{0}$ and $D^{+}D^{-}$  around the mass range of 
$G(3900)$, $\psi(4040)$, $\psi(4160)$, $Y(4260)$, and $\psi(4415)$, etc., are foreseen. 
These  results offer crucial experimental insights into the nature of hadron production in the open-charm region.
\end{abstract}

\maketitle
%\footnote{Full author list given at the end of the Letter}
The production of hadrons in $\EE$ annihilation above the open-charm threshold is a topic of ongoing theoretical and experimental research.
In 1980, a theoretical calculation for the charm cross section in $\EE$ annihilation was first attempted based on a coupled-channel potential model~\cite{Eichten:1979ms}.
This calculation presented a prediction of the $\Delta R$ value ($\Delta R = \sum_{i}R_{i}$, where $R_{i}$ stands for the ratio of individual hadron cross section to muon cross section in electron-positron collisions $i$ runs over the two-body channels) with the $D\bar{D}$ final states.
According to this prediction, there are five vector charmonium states between 3.773 GeV [$\psi(3773)$, $1D$ state] and 4.95~GeV, namely, the $3S$, $2D$, $4S$, $3D$, and $5S$ states, dominated by the $D\bar{D}$ final states.
In experimental studies, besides the three well-established structures observed in the inclusive hadronic cross sections~\cite{PDG2020}, {\it i.e.}, $\psi(4040)$, $\psi(4160)$, and $\psi(4415)$, many new states, such as $Y(4230)$, $Y(4260)$, $Y(4360)$, $Y(4660)$, have been reported in the initial state radiation (ISR) processes at the  $B$ factories~\cite{Y4260Y4360_barbar, Y4260Y4360_barbar00, Y4260Y4360_barbar01, Y4260Y4360_barbar02, Y4260Y4360_belle,Y4260Y4360_belle00, Y4260Y4360_belle01,Y4260Y4360_belle01,Y4260Y4360_belle02,Y4260Y4360_belle03} or in the direct $e^+e^-$ production at the CLEO~\cite{CLEO} and BESIII experiments~\cite{BESIIIAB,BESIIIAC,BESIIIAD,BESIIIcuv,BESIIIccp, BESIII:2023rse, BESIII:2024umc, BESIII:2024ogz}.
Among them, the BESIII experiment  found that the mass of $Y(4360)$ is around 4.3 GeV/$c^{2}$~\cite{BESIIIAC}, which is different from the value given by the Particle Data Group (PDG)~\cite{PDG2020}.
The overpopulation of structures in this region and the mismatch of the properties between the potential model predictions and experimental measurements have led to various interpretations, such as hybrid states, tetraquark states, or molecular states~\cite{QCD, Brambilla:2019esw}.
Although this information enriches our understanding of these exotic structures, the nature of these states is still not understood.

The studies of charmed meson pairs in $\EE$ annihilation above the open-charm threshold  are expected to clarify the current understanding of these states. 
At present, the available observed cross sections of the $\EE\ar D\bar{D}$ process with limited energy points have been reported by $B$ factories~\cite{Aubert:2008pa, Pakhlova:2008zza} using  the ISR process and through direct $e^+e^-$ production at the CLEO experiment~\cite{CroninHennessy:2008yi}. 
Although the interpretations for the possible structure featured in the $D\bar{D}$ final states are performed~\cite{Zhang:2009gy, Du:2016qcr}, the understanding for the properties of vector charmonium(like) states is still limited except for the $\psi(3770)$ state.
A precise measurement, particularly of the exclusive Born cross sections for $\EE\ar D\bar{D}$, is highly desirable to validate the interpretations of the established states and provide insight into the energy region above the open-charm threshold.

In this Letter, we report a precise measurement of Born cross sections for the $\EE\ar D^{0}\bar{D}^{0}$ and $\EE\ar  D^{+}D^{-}$ processes, specifically, at 150 center-of-mass (c.m.) energy points.
Many clear structures in the line shape of $\EE\ar D^{0}\bar{D}^{0}$ and $\EE\ar  D^{+}D^{-}$ in the c.m. energy at 3.90, 4.05, 4.20, 4.42 GeV, etc., can be seen.
The datasets used in this work correspond to a total luminosity of approximately 20 fb$^{-1}$ of $\EE$ collisions, which includes the so-called $XYZ$ data sample~\cite{Ablikim:2015nan,BESIII:2022dxl} and the $R$-scan data sample~\cite{BESIII:2017lkp}, collected at c.m. energies from 3.80 to 4.95 GeV with the BESIII detector~\cite{Wang:2007tv} at the BEPCII collider~\cite{BESIII}.

The BESIII detector's cylindrical core, encapsulating \(93\%\) of the \(4\pi\) solid angle, integrates a helium-based multilayer drift chamber (MDC), a plastic scintillator time-of-flight (TOF) system, and a CsI (Tl) electromagnetic calorimeter. These components are all nestled within a superconducting solenoidal magnet that generates a \(1.0~\text{T}\) magnetic field. An octagonal flux-return yoke, fortified with steel-interleaved resistive plate chamber muon identifier modules, supports the solenoid. This sophisticated setup achieves a charged-particle momentum resolution of \(0.5\%\) at \(1~\text{GeV}/c\), and a \(dE/dx\) resolution of \(6\%\) for electrons emanating from Bhabha scattering. The TOF system's barrel section boasts a time resolution of \(68~\text{ps}\), in contrast to the end cap section's \(110~\text{ps}\). Notably, the end cap TOF system underwent an enhancement in 2015, adopting multigap resistive plate chamber technology to furnish a time resolution of \(60~\text{ps}\).

In order to achieve a high efficiency for the selection of $\EE\ar D\bar{D}$ events,  we employ a single tag technique instead of a full reconstruction.
With this technique we reconstruct only one $D^{0}$($D^{+}$) meson through the $K^{-}\pi^{+}\pi^{+}\pi^{-}$($K^{-}\pi^{+}\pi^{+}$) mode, while the corresponding antiparticle $\bar{D}^{0}$($D^{-}$) is extracted from the recoil side. 
Unless otherwise noted, the charge-conjugate  mode of the $D^{0}$($D^{+}$) process is included by default.  
To determine the detection efficiency for $\EE\ar D\bar{D}$, 50,000 simulated events are generated for each energy point using the \textsc{kkmc} generator~\cite{kkmc,Jadach:2000ir}  according to the  VSS model~\cite{evt1,evt2}, where the ISR effect is included.
The $D^{0}$($D^{+}$) meson decays to the $K^{-}\pi^{+}\pi^{+}\pi^{-}$($K^{-}\pi^{+}\pi^{+}$) modes are simulated with the amplitude sampling~\cite{D0tokpipipi} via \textsc{evtgen}~\cite{evt1,evt2}, and the antimesons are set to decay inclusively according to the known branching fractions provided by PDG~\cite{PDG2020}. 
The response of the BESIII detector is modeled with Monte Carlo (MC) simulation using a framework based on \textsc{geant}{\footnotesize 4}~\cite{geant4,Allison:2006ve}.

Charged tracks are reconstructed in the MDC with points of closest approach to the $\EE$ interaction point that are within 10 cm in the beam direction and 
1~cm transverse to the beam direction and within the angular coverage of the MDC $|\cos\theta|<0.93$, where $\theta$ is the polar angle with respect to the symmetry axis of the MDC.
Information from the specific ionization energy loss   measured in the MDC, combined with the time of flight,  is used to determine the particle identification (PID) confidence levels for the pion and kaon hypotheses. 
Each track is assigned to the particle type with the higher probability. 
For the $D^{0}$ mode, events with at least one negatively charged kaon, one negatively charged pion, and two positively charged pions are kept for further analysis, for the $D^{+}$ mode, events with at least one negatively charged kaon and two positively charged pions for the $D^{+}$ mode, are kept for further analysis.
The $D^{0}$($D^{+}$) candidates are reconstructed from the $K^{-}\pi^{+}\pi^{+}\pi^{-}$($K^{-}\pi^{+}\pi^{+}$) combination by requiring its invariant mass to be within 14~(16)~MeV/$c^{2}$ around the nominal $D^{0}$($D^{+}$) mass, which corresponds to 3 times  the mass resolution.

\begin{figure}[!htbp]
\centering
\includegraphics[width=0.45\textwidth]{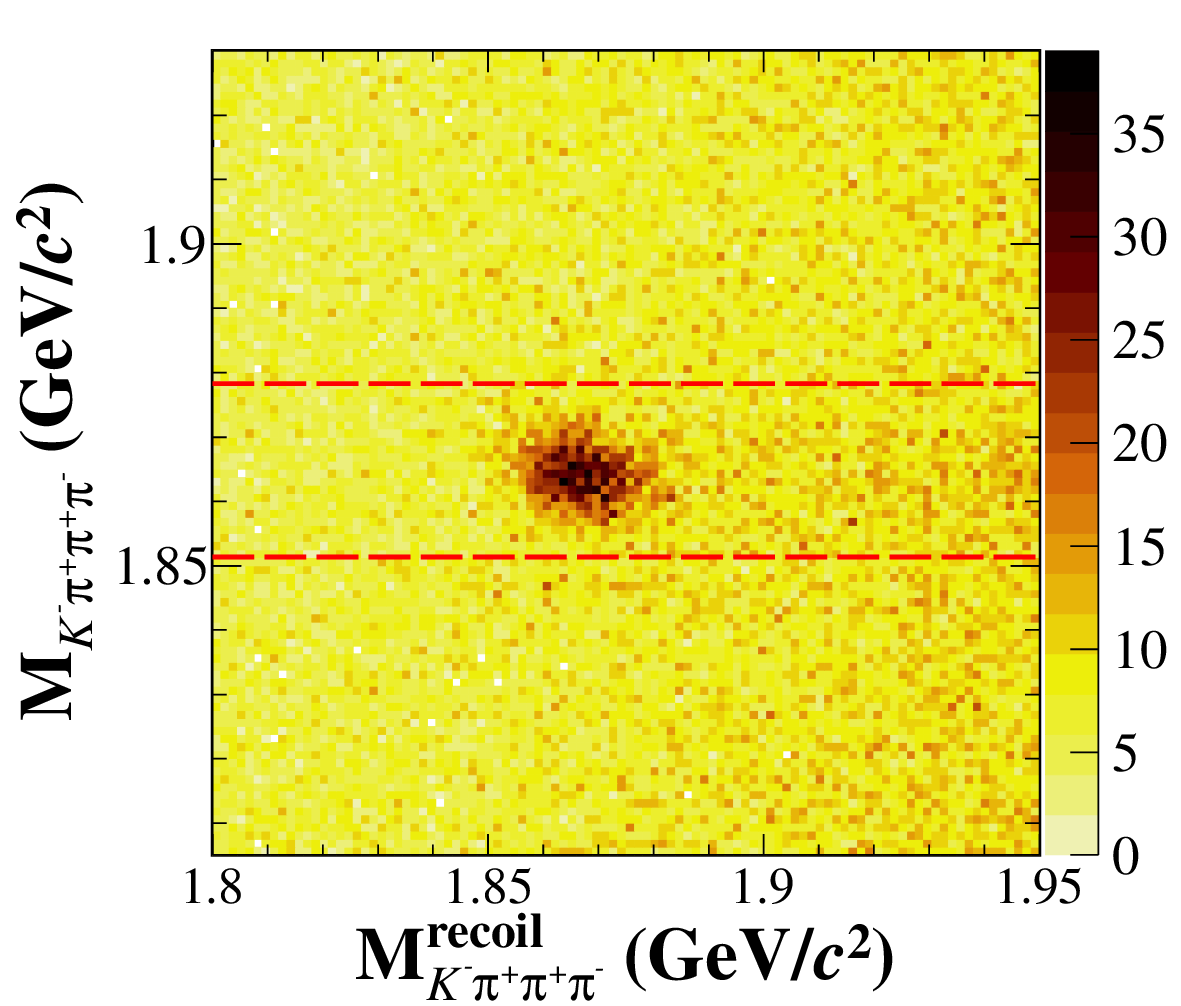}
\includegraphics[width=0.45\textwidth]{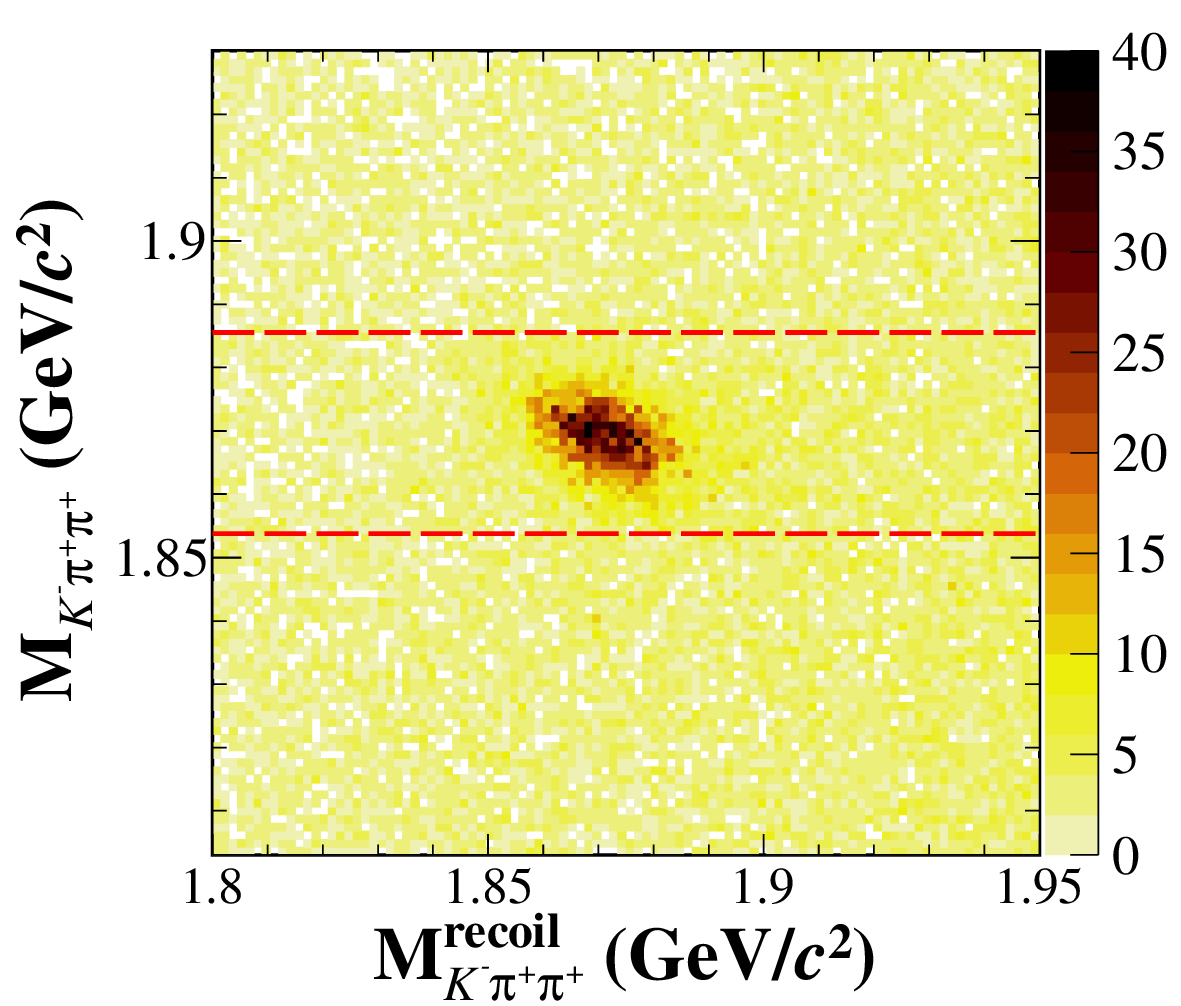}
\caption{The two-dimensional distributions of $M_{D}$ versus $M^{\rm recoil}_{D}$ at $\sqrt{s} = 4.1992$ GeV. The dashed lines represent the signal region for the $D$ meson in the $D^{0}\bar{D}^{0}$ mode (top) and the $D^{+}D^{-}$ mode (bottom).}
\label{scatterplot}
\end{figure}

The antimeson candidates $\bar{D}^{0}$($D^{-}$) are inferred by the mass recoiling against the tagged meson ($M^{\rm recoil}_{D}$) via the $K^{-}\pi^{+}\pi^{+}\pi^{-}$($K^{-}\pi^{+}\pi^{+}$) system:
\begin{equation}
M^{\rm recoil}_{D} = \sqrt{(\sqrt{s}-E_{D})^{2}- |\textbf{p}_{D}|^{2}},
\end{equation}
where $E_{D}$ and $\textbf{p}_{D}$ are the energy and momentum of the selected $K^{-}\pi^{+}\pi^{+}\pi^{-}$($K^{-}\pi^{+}\pi^{+}$) candidate in the c.m. system, respectively, and $\sqrt{s}$ is the c.m. energy~\cite{BESIII:2020eyu}.
To improve the resolution, a correction is applied to $M^{\rm recoil}_{D}$ given by $M^{\rm recoil}_{D} + M_{D} - m_{D}$,  where $M_{D}$ is the invariant mass of the selected $K^{-}\pi^{+}\pi^{+}\pi^{-}$($K^{-}\pi^{+}\pi^{+}$) candidate and $m_{D}$ is the nominal $D$ mass~\cite{PDG2020}. Figure~\ref{scatterplot} shows the 2D distributions of $M^{\rm recoil}_{D}$ versus $M_{D}$ for the $K^{-}\pi^{+}\pi^{+}\pi^{-}$($K^{-}\pi^{+}\pi^{+}$) final states at $\sqrt{s} = 4.1992$ GeV.
After applying all the aforementioned selection criteria,   the remaining background exhibits a smooth shape in the region of interest based on the sideband study.

\begin{figure}[!htbp]
\centering
\includegraphics[width=0.45\textwidth]{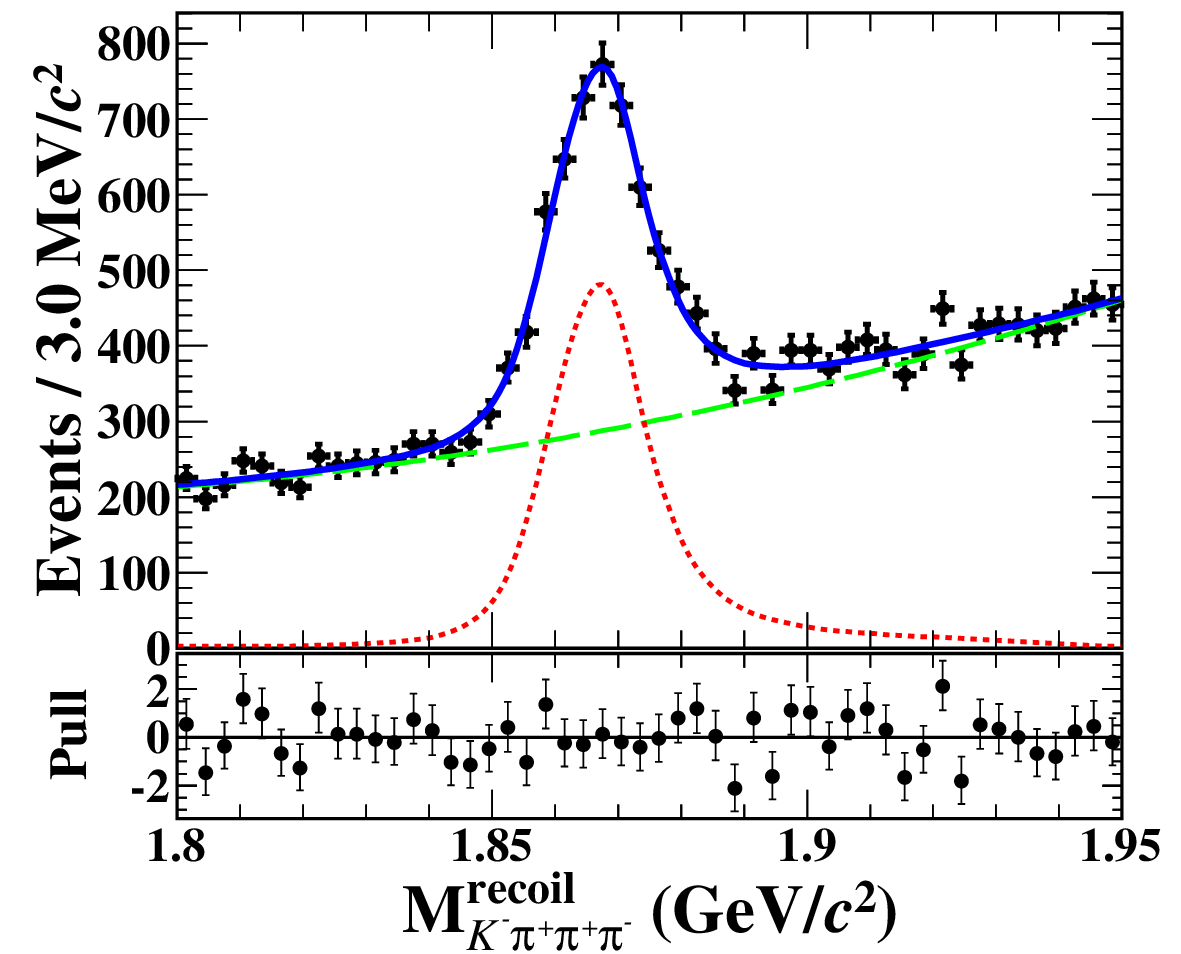}
\includegraphics[width=0.45\textwidth]{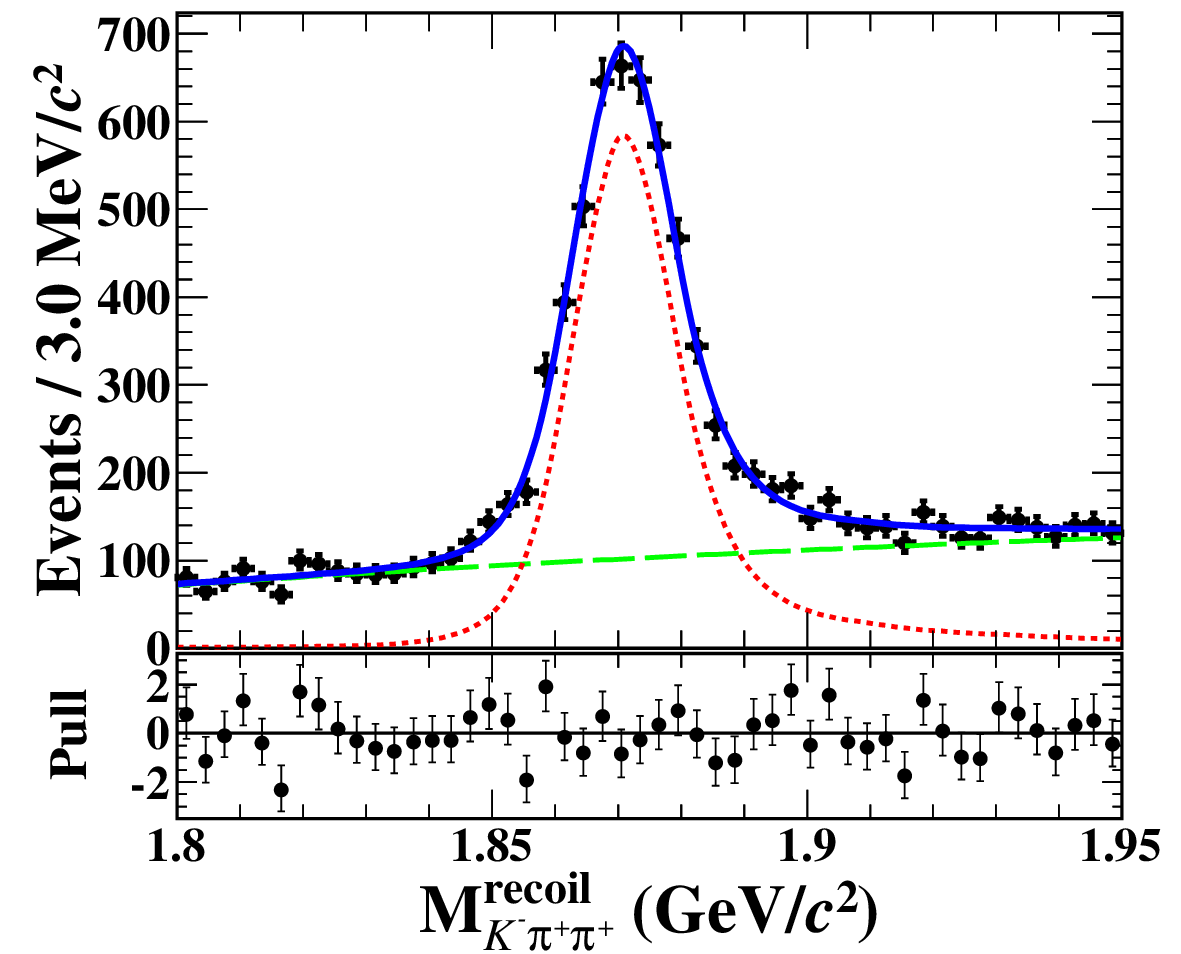}
\caption{Fits to the $M^{\rm recoil}_{D}$ spectra at $\sqrt{s} = 4.1992$ GeV for the $D^{0}\bar{D}^{0}$ mode (top) and the $D^{+}D^{-}$ mode (bottom).
The dots with error bars represent the data, the blue solid lines indicate the fit results, the red short-dashed lines represent the signal, and the green long-dashed lines correspond to the background.}
\label{recoil_fitting}
\end{figure}

The signal yields for the $\EE\to D\bar{D}$ process at each energy point are extracted by performing an extended maximum likelihood fit to the $M^{\rm recoil}_{D}$ spectrum in the range from 1.80 to 1.95 GeV/$c^{2}$. 
In the fit, the signal shape for the $\EE\ar D\bar{D}$ process is described by the convolution of the MC-simulated shape with a Gaussian function, which accounts for the difference in mass resolution between the data and the MC simulation. 
The parameters of the Gaussian function are floating for $XYZ$ data points and  some $R$-scan data points with higher statistics, while they  are fixed for the other $R$-scan data points with low statistics.
The fixed parameters are determined from the fits to the neighbor  $XYZ$ data points.
The background contributions  are modeled using a second-order polynomial function.
Figure~\ref{recoil_fitting} illustrates the results of the fits  to the $M^{\rm recoil}_{D}$ distributions at $\sqrt{s} = 4.1992$ GeV.
The signal yields obtained from the fits are summarized in Supplemental Material at~\cite{ABC}.

\begin{figure*}[!htbp]
\begin{center}
\includegraphics[width=1.0\textwidth]{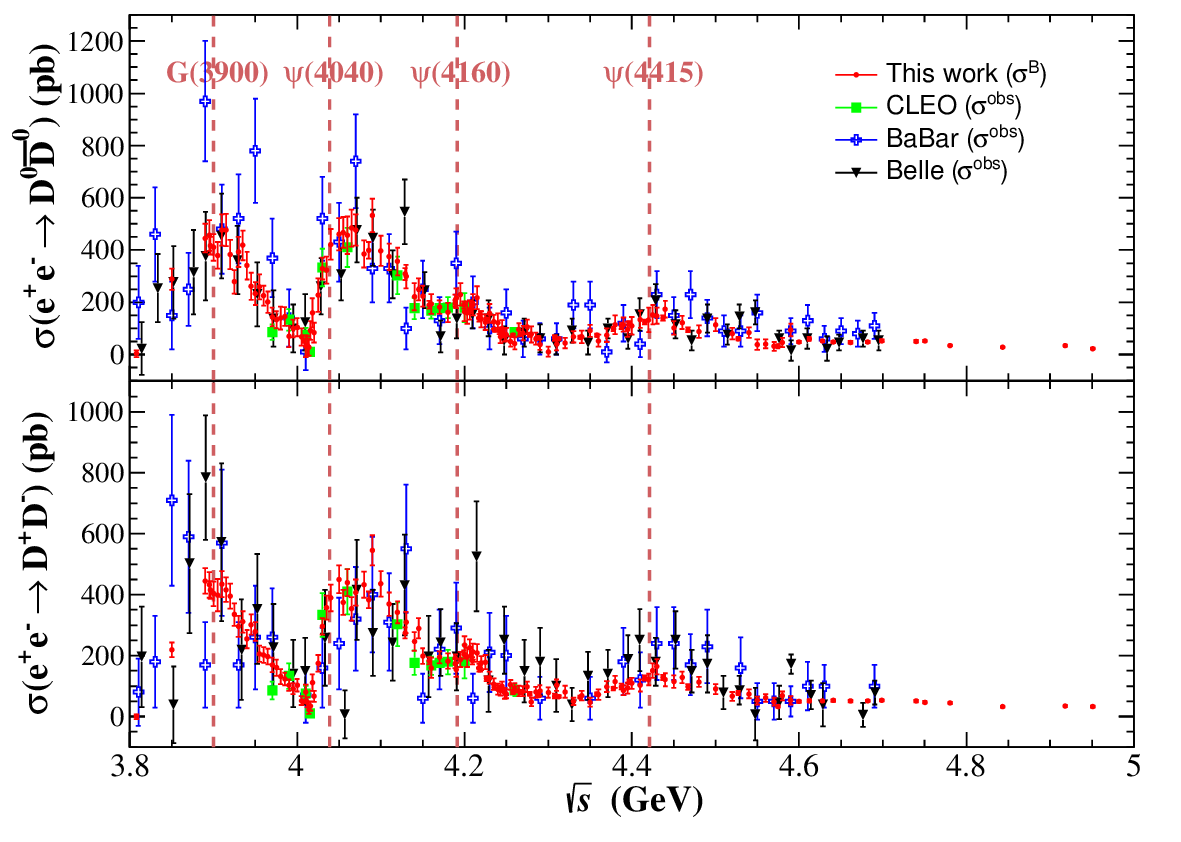}
\caption{Cross sections with total uncertainties for the $\EE\ar D\bar{D}$ process as a function of c.m. energy from 3.80 to 4.95 GeV compared between this work for Born cross section ($\sigma^B$) and the previous measurements for observed cross section ($\sigma^{obs}$). Clear peaks around the mass range of $G(3900)$, $\psi(4040)$, $\psi(4160)$, $\psi(4415)$, etc. can be seen.
}
\label{scatterplot:BC}
\end{center}
\end{figure*}

The Born cross section for $\EE\to D\bar{D}$ is calculated by
\begin{equation}
\sigma^{B}(s) =\frac{N_{\rm obs}}{2{\cal{L}}(1 + \delta)\frac{1}{|1 - \prod|^{2}}\epsilon{\cal B}},
\end{equation}
where $N_{\rm obs}$ represents the number of  observed signal events,
the factor of 2 accounts for the charge-conjugate mode,
${\cal{L}}$ corresponds to the integrated luminosity, $(1 + \delta)$ is the ISR correction factor, $\frac{1}{|1 - \prod|^{2}}$ is the vacuum polarization correction factor~\cite{vppre}, $\epsilon$ denotes the detection efficiency, and $ {\cal B}$ stands for the  branching fractions of $D^{0}\to K^{-}\pi^{+}\pi^{+}\pi^{-}$ mode and $D^{+}\to K^{-}\pi^{+}\pi^{+}$ mode, taken from the Particle Data Group~\cite{PDG2020}.
The ISR correction factor is obtained through QED calculations~\cite{Kuraev:1985hb}, where the cross sections measured in this analysis are used as inputs and iterated until convergence. The measured Born cross sections, along with the results from the CLEO-c~\cite{CroninHennessy:2008yi}, {\it{BABAR}}~\cite{Aubert:2008pa}, and Belle~\cite{Pakhlova:2008zza} experiments, are shown in Fig.~\ref{scatterplot:BC} and summarized in Ref.~\cite{ABC}, which also includes all the numbers used in the calculation.
To evaluate the resonance influence to the measurement of Born cross section,
a least-$\chi^{2}$ method combined with the simultaneous fit of the dressed cross sections $\sigma^{\rm dressed} = \sigma^{B}/|1 - \prod|^{2}$ for the $\EE\to D^0\bar{D}^0$ and $D^+D^-$ processes, parametrized as the coherent sum of eight relativistic Breit-Wigner ($BW$) functions:
\begin{equation}
\sigma^{\rm dressed}(\sqrt{s}) = \left|\sum_{k=1}^{8}e^{i\phi_{k}}BW_{k}(\sqrt{s})\sqrt{\frac{P(\sqrt{s})}{P(M)}}\right|^{2},
\end{equation}
has been attempted,
where $BW(\sqrt{s})$ is given by
\begin{equation}
BW(\sqrt{s}) = \frac{\sqrt{12\pi\Gamma_{ee}{\cal{B}}\Gamma}}{s - M^{2} + iM\Gamma}.
\end{equation}
Here, the masses $M$ and total widths $\Gamma$ for seven known resonances $\psi(3770)$, $\psi(4040)$, $\psi(4160)$, $Y(4230)$, $Y(4360)$, $\psi(4415)$, and $Y(4660)$ are fixed at individual PDG values~\cite{PDG2020}, while they are free for another structure regarded as $G(3900)$ around 3.9  GeV~\cite{Lin:2024qcq, Husken:2024hmi, Salnikov:2024wah}.
For $\Gamma_{ee}$,  electronic partial widths, and $\cal{B}$, the branching fractions of the decay, they are free for all resonances.
The relative phase between different $BW$ functions is denoted by $\phi$, which is set to be  different in the simultaneous fit for both modes. And $P(\sqrt{s})$ represents the two-body phase space factor. To account for the beam energy spread, $\sigma^{\rm dressed}(\sqrt{s})$ is convolved with a Gaussian function with the standard deviation $\sigma = (-2.147 + 0.9454\sqrt{s})$ GeV. 
The influence to the measurement of Born cross section has been incorporated in the uncertainty of the line-shape model below.
Note that the parameters for all assumed resonances strongly depend on the chosen fit model and indicating the need for further in-depth research, such as a coupled-channel K-matrix analysis~\cite{Husken:2024hmi}.
The details for this fit  are listed in Supplemental Material~\cite{ABC}.

Systematic uncertainties in the measurement of the cross sections for the $\EE\rightarrow D^{0}\bar{D}^{0}$ and $D^{+}D^{-}$ processes originate from the luminosity measurement, the efficiencies of tracking and PID, the requirement of the $D$ mass window, the fit of the $M^{\rm recoil}_{D}$ spectrum, the branching fractions of $D^{0}\rightarrow K^{-}\pi^{+}\pi^{+}\pi^{-}$ and $D^{+}\rightarrow K^{-}\pi^{+}\pi^{+}$, and the line-shape structures.
The uncertainty due to the vacuum polarization is negligible. 
The integral luminosity is measured to an uncertainty up to 1.0\%~\cite{Ablikim:2015nan}.
The uncertainties due to data-MC differences in BESIII tracking and particle ID efficiencies are both 1\% per track~\cite{pid}.
The requirement of the $D$ mass window is studied by varying the nominal requirements by $\pm1\sigma$, resulting in uncertainties of 2.1\% and 1.8\% in the neutral and charged modes, respectively.
The systematic uncertainty arising from the fit of the $M^{\rm recoil}_{D}$ spectrum includes the fit range and the background shape. Varying the mass range by $\pm$ 5 MeV/$c^{2}$ results in an uncertainty of 1.5\% for the neutral mode and 1.6\% for the charged mode.
The uncertainty due to the background shape is estimated to be 1.0\% for the neutral mode and 0.9\% for the charged mode using alternative fits with a second- or third-order polynomial function.
The branching fractions of $D^{0}\rightarrow K^{-}\pi^{+}\pi^{+}\pi^{-}$ and $D^{+}\rightarrow K^{-}\pi^{+}\pi^{+}$ are quoted with uncertainties of 2.4\% for both modes from the PDG~\cite{PDG2020}.
The uncertainty arising from the line-shape model, including the ISR correction factor, is estimated by comparing the $(1 + \delta) \cdot \epsilon$ values  with and without the addition of one more resonance in the fit of the cross sections as the input line shape, which introduces an uncertainty of 1.5\% for the neutral mode and 1.6\% for the charged mode between the nominal and alternative models.
Assuming all sources are independent, the total systematic uncertainties on the cross section measurements are determined to be 7.0\% for the $\EE\rightarrow D^{0}\bar{D}^{0}$ mode and 6.5\% for the $\EE\rightarrow D^{+}D^{-}$ mode by quadratic sum.

In summary, a measurement of exclusive Born cross sections for the $\EE\rightarrow D^{0}\bar{D}^{0}$ and $D^{+}\bar{D}^{-}$ processes is presented at 150 c.m. energy points ranging from 3.80 to 4.95~GeV with unprecedented precision. 
The result is in qualitative agreement with previous experiments~\cite{Aubert:2008pa, Pakhlova:2008zza, CroninHennessy:2008yi} and the prediction of the coupled-channel model~\cite{Eichten:1979ms}.
Many clear peaks  in the line shape of $\EE\rightarrow D\bar{D}$ around the mass range of 
$G(3900)$, $\psi(4040)$, $\psi(4160)$, $\psi(4260)$, and $\psi(4415)$, etc. are identified.
It implies that there may be some potential contributions from charmonium(like) states. In particular, the possible structure around 3.9 GeV 
which was featured and interpreted by $G(3900)$ at the $B$ factories~\cite{Aubert:2008pa, Pakhlova:2008zza},
has been discussed recently by the theoretical models  as the first $P$-wave $D\bar{D}^{*}$ molecular resonance~\cite{Lin:2024qcq},
 threshold enhancement~\cite{Husken:2024hmi} or the final-state interaction~\cite{Salnikov:2024wah}. Thus, more detailed study related to a coupled-channel K-matrix analysis is needed to validate this structure.

Our results for describing the cross section provided in Supplemental Material depend on the chosen model, which simplifies the analysis by ignoring the interactions between different decay channels. However, according to the model calculation by the Cornell group~\cite{Eichten:1979ms}, strong coupled-channel effects need to be considered, which also is proposed by the recent theoretical works~\cite{Lin:2024qcq, Husken:2024hmi, Salnikov:2024wah}.
It is out of the scope of this work, but a more comprehensive approach based on K-matrix formalism to fit the cross section results of various exclusive channels  
 is expected to test the scenarios ~\cite{Wang:2019mhs,Wang:2020prx,2005OZI,2009OZI,2021OZI, Zhang:2009gy, Du:2016qcr} of charmonium(-like) states  above the open-charm threshold. This work provides important experimental evidences with unprecedented precision and insights into the nature above the open-charm region, especially for  the understanding of the property of charmoniumlike states.

{\it{Acknowledgement}}---The BESIII Collaboration thanks the staff of BEPCII and the IHEP computing center for their strong support. This work is supported in part by National Key R\&D Program of China under Contracts No. 2020YFA0406300, No. 2020YFA0406400; National Natural Science Foundation of China (NSFC) under Contracts No. 12075107, No. 12247101, No. 11635010, No. 11735014, No. 11835012, No. 11935015, No. 11935016, No. 11935018, No. 11961141012, No. 12022510, No. 12025502, No. 12035009, No. 12035013, No. 12061131003, No. 12192260, No. 12192261, No. 12192262, No. 12192263, No. 12192264, No. 12192265, No. 12221005, No. 12225509, No. 12235017; the 111 Project under Grant No. B20063; the Chinese Academy of Sciences (CAS) Large-Scale Scientific Facility Program; the CAS Center for Excellence in Particle Physics (CCEPP); Joint Large-Scale Scientific Facility Funds of the NSFC and CAS under Contract No. U1832207; CAS Key Research Program of Frontier Sciences under Contracts No. QYZDJ-SSW-SLH003, No. QYZDJ-SSW-SLH040; 100 Talents Program of CAS; The Institute of Nuclear and Particle Physics (INPAC) and Shanghai Key Laboratory for Particle Physics and Cosmology; European Union's Horizon 2020 research and innovation programme under Marie Sklodowska-Curie grant agreement under Contract No. 894790; German Research Foundation DFG under Contracts No. 455635585, Collaborative Research Center CRC 1044, FOR5327, and GRK 2149; Istituto Nazionale di Fisica Nucleare, Italy; Ministry of Development of Turkey under Contract No. DPT2006K-120470; National Research Foundation of Korea under Contract No. NRF-2022R1A2C1092335; National Science and Technology fund of Mongolia; National Science Research and Innovation Fund (NSRF) via the Program Management Unit for Human Resources and Institutional Development, Research and Innovation of Thailand under Contract No. B16F640076; Polish National Science Centre under Contract No. 2019/35/O/ST2/02907; The Swedish Research Council; and U. S. Department of Energy under Contract No. DE-FG02-05ER41374.

%\newpage
\begin{widetext}
\begin{center}
%\onecolumngrid
%\noindent
\small
M.~Ablikim$^{1}$, M.~N.~Achasov$^{5,b}$, P.~Adlarson$^{75}$, X.~C.~Ai$^{81}$, R.~Aliberti$^{36}$, A.~Amoroso$^{74A,74C}$, M.~R.~An$^{40}$, Q.~An$^{71,58}$, Y.~Bai$^{57}$, O.~Bakina$^{37}$
% ...
\\
\vspace{0.2cm}

(BESIII Collaboration)

%\vspace{0.2cm} 
{\it
$^{1}$ Institute of High Energy Physics, Beijing 100049, People's Republic of China\\
% ...
$^{5}$ Budker Institute of Nuclear Physics SB RAS (BINP), Novosibirsk 630090, Russia\\
% ...
$^{36}$ Johannes Gutenberg University of Mainz, Johann-Joachim-Becher-Weg 45, D-55099 Mainz, Germany\\
$^{37}$ Joint Institute for Nuclear Research, 141980 Dubna, Moscow region, Russia\\
% ...
$^{40}$ Liaoning Normal University, Dalian 116029, People's Republic of China\\
% ...
$^{57}$ Southeast University, Nanjing 211100, People's Republic of China\\
$^{58}$ State Key Laboratory of Particle Detection and Electronics, Beijing 100049, Hefei 230026, People's Republic of China\\
% ...
$^{71}$ University of Science and Technology of China, Hefei 230026, People's Republic of China\\
% ...
$^{74}$ University of Turin and INFN, (A)University of Turin, I-10125, Turin, Italy; (B)University of Eastern Piedmont, I-15121, Alessandria, Italy; (C)INFN, I-10125, Turin, Italy\\
$^{75}$ Uppsala University, Box 516, SE-75120 Uppsala, Sweden\\
% ...
$^{81}$ Zhengzhou University, Zhengzhou 450001, People's Republic of China\\
\vspace{0.2cm}
% ...
$^{b}$ Also at the Novosibirsk State University, Novosibirsk, 630090, Russia\\
% ...
}
\end{center}
\end{widetext}
 
\end{document}